\documentclass[useAMS,usenatbib]{mn2e}
\usepackage{epsfig}

\def\lesssim{\mathrel{\hbox{\rlap{\hbox{\lower4pt\hbox{$\sim$}}}\hbox{$<$}}}}
\def\gtrsim{\mathrel{\hbox{\rlap{\hbox{\lower4pt\hbox{$\sim$}}}\hbox{$>$}}}}
\def\la{\mathrel{\hbox{\rlap{\hbox{\lower4pt\hbox{$\sim$}}}\hbox{$<$}}}}
\def\ga{\mathrel{\hbox{\rlap{\hbox{\lower4pt\hbox{$\sim$}}}\hbox{$>$}}}}

\def\spose#1{\hbox to 0pt{#1\hss}}
\def\approxlt{\mathrel{\spose{\lower 3pt\hbox{$\sim$}}
	\raise 2.0pt\hbox{$<$}}}
\def\approxgt{\mathrel{\spose{\lower 3pt\hbox{$\sim$}}
	\raise 2.0pt\hbox{$>$}}}

\def\<{\thinspace}
\def\s{\hbox{\phantom{5}}}	
\def\ss{\s\s}		

\def\boxit#1{\vbox{\hrule\hbox{\vrule\kern3pt\vbox{\kern3pt
          #1 \kern3pt}\kern3pt\vrule}\hrule}}

\def\deg{^o}

\def\ga{{\rm\thinspace gauss}}

\def\s{{\rm\thinspace s}}

\def\h50{\hbox{$\rm\thinspace h_{50}$}}
\def\h50m1{\hbox{$\rm\thinspace h_{50}^{-1}$}}

\title[Three newly discovered globular clusters in NGC 6822] {Three
  newly discovered globular clusters in NGC 6822}
\author[Huxor et al.]{A. P. Huxor$^{1}$\thanks{Email:avon@ari.uni-heidelberg.de}, A. M. N. Ferguson$^{2}$,    J. Veljanoski$^{2}$, A. D. Mackey$^{3}$, N. R. Tanvir$^{4}$\\
  $^{1}$Astronomisches Rechen-Institut, Zentrum f\"{u}r Astronomie der Universit\"{a}t Heidelberg, M\"{o}nchstra{\ss}e 12 - 14, \\
  69120 Heidelberg, Germany.\\
  $^{2}$SUPA, Institute for Astronomy, University of Edinburgh, Royal Observatory, Blackford Hill, Edinburgh EH9 3HJ, UK\\
  $^{3}$Research School of Astronomy \& Astrophysics, Australian National University, Mt. Stromlo Observatory, Cotter Road, \\
  Weston Creek, ACT 2611, Australia \\
  $^{4}$Department of Physics and Astronomy, University of Leicester, University Road, Leicester LE1 7RH, UK\\
}
\begin{document}

\date{}

\pagerange{\pageref{firstpage}--\pageref{lastpage}} \pubyear{2007}

\maketitle

\label{firstpage}

\begin{abstract}

  We present three newly discovered globular clusters (GCs) in the
  Local Group dwarf irregular NGC 6822. Two are luminous and compact,
  while the third is a very low luminosity diffuse cluster. We report
  the integrated optical photometry of the clusters, drawing on
  archival CFHT/Megacam data.  The spatial positions of the new GCs
  are consistent with the linear alignment of the already-known
  clusters. The most luminous of the new GCs is also highly
  elliptical, which we speculate may be due to the low tidal field in
  its environment.

\end{abstract}

\begin{keywords}
galaxies: star clusters -- galaxies: individual (NGC 6822) 
\end{keywords}

\section{Introduction}

In $\Lambda$CDM cosmology, large galaxies are assembled as the result
of the accretion and merger of smaller galaxies. If the globular
cluster (GC) systems of large galaxies are also formed, at least in
part, from the accretion of GCs from the smaller systems, then the GCs
themselves can act as beacons of this process. Indeed, the seminal
study of the Galactic GCs by \citet{SearleZinn78} was crucial for
understanding the history of our own Milky Way (MW). Being compact and
luminous, GCs are excellent probes when field star populations cannot
be resolved. In our previous work \citep{Huxoretal11,Mackeyetal10}, we
have shown that the GC population of the outer halo of M31 arises
largely from the accretion of dwarf galaxies. In many cases, one can
identify the remnants of dwarf galaxies in the process of delivering
their clusters into M31, where they are becoming part of its retinue
of GCs. A similar process is also taking place in the MW, where the
Sagittarius dwarf is most likely contributing its GCs to the MW halo
\citep{Bellazzinietal03, law10}. Recent work by \citet{Kelleretal12}
also supports the view that many GCs found in the outer halo of the MW
have been accreted alongside their (now disrupted) dwarf galaxy hosts.
They conclude that the MW halo has experienced the accretion of some
three Magellanic-like or equally up to 30 Sculptor-like dwarf
galaxies, or some intermediate mix of both types.  Knowledge of the
characteristics of the types of GCs found in a range of dwarf galaxies
will assist in determining which scenario may have occurred.

To use GCs as probes, it is essential to understand the properties of
the GC systems of dwarf galaxies, the relationship of these properties
to their host galaxies, and whether we can still identify these after
they have been accreted into a more massive galaxy. Dwarf irregulars
are particularly interesting in this regard as they are usually found
in the field. Their relative isolation makes them ideal laboratories
for studying the pristine properties GC systems. This contrasts with
dwarf spheroidal and elliptical galaxies which are usually found close
to more massive galaxies, and thus they (and their GCs) will likely
have been influenced by them. Motivated by our previous work in which
wide-area searches yielded the discovery of many new clusters in M31
\citep{Huxoretal08, Martinetal06}, M33 \citep{Huxoretal09,
  Cockcroftetal11}, and the M31 satellite galaxies NGC 185 and NGC 147
(Veljanoski et al. in prep), we decided to investigate the outer
regions of the dwarf irregular NGC 6822 which benefits from extensive
archival CFHT/MegaCam imaging, and in which \citet{Hwangetal11} have
recently discovered four new extended star clusters.
  
\begin{figure}
 \centering
 \includegraphics[angle=0,width=70mm]{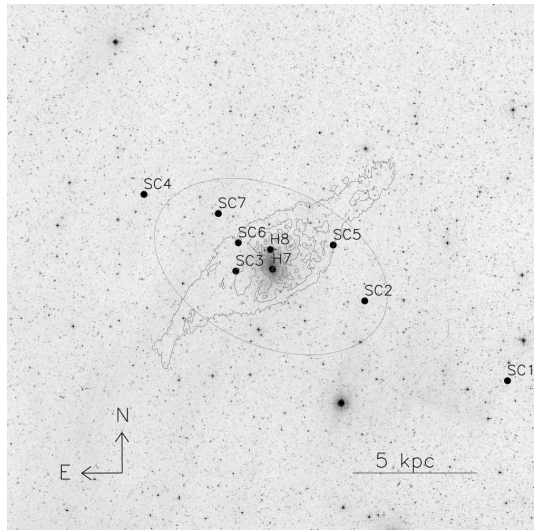}
 \vspace{2pt}
 \caption{Image from the DSS of NGC 6822 with locations of the new and
   previously-known clusters shown, in filter GSS bandpass number 36.
   The image is 2.57 x 2.57 degrees. The ellipse shows the extent of
   the RGB stars of the spheroid where the halo is detected above the
   noise \citep{Battinellietal06}, with a semi-major axis of 36\arcmin
   and an ellipticity of 0.36.  Also shown are contours from the HI
   map of \citet{deBlokWalter00}.}\label{Fi:dss}
\end{figure}

NGC 6822 is a member of the Local Group, and is not associated with
either the MW or M31. We use an adopted distance of 472 kpc determined
as the average of published values for which the error of the distance
modulus is $<$ 0.2 mags \citep{Gorskietal11}. It has an absolute
magnitude $M_{V}$ of --15.2 \citep{Mateo98} and an R$_{25}$ of 465
arcsec (RC3.9 value, reported by
NED\footnote{http://ned.ipac.caltech.edu/}) equal to $\sim$ 1 kpc at
our adopted distance. NGC 6822 possesses a number of interesting
features including a ring of gas and stars which is almost
perpendicular to the main body of the galaxy \citep{deBlokWalter00}.
The galaxy has also been found to have an extended stellar spheroid
\citep{Battinellietal06} (see Figure \ref{Fi:dss}). The central
regions of NGC 6822 contain many young massive clusters
\citep{Chandaretal00}; however, until very recently, it was believed
that there was only one truly old globular cluster in the system
\citep{Grebel02}, known as Hubble VII -- one of his original list of
``nebulae"\footnote{The majority of these are HII regions.  Hubble VI
  is a young cluster and the nature of Hubble IX is still unclear.} in
NGC 6822 \citep{Hubble25}.  \citet{CohenBlakeslee98} undertook a
spectroscopic study of this object, reporting an age of ~11 Gyr and
[Fe/H] = --1.95 dex. Another of Hubble's candidates, Hubble VIII, has
also been subjected to detailed study and appears to be a massive
intermediate-age cluster. Using HST/WFPC2 observations,
\citet{Wyderetal00} derived an age of ~1.5 Gyr however a spectroscopic
study by \citet{Straderetal03} found it to be somewhat older at 3--4
Gyr. Both \citet{Chandaretal00} and \citet{CohenBlakeslee98} derive
spectroscopic [Fe/H] estimates for Hubble VIII, finding a value of
about --2.0 dex.

In addition, NGC 6822 possesses four extended star clusters (SC1--SC4,
shown in Figure \ref{Fi:dss}) that have been found beyond the main
body of the galaxy \citep{Hwangetal11}.  These were discovered in a
wide-field CFHT/Megacam survey of NGC 6822 that covered a region of
3$^{\circ}$ $\times$ 3$^{\circ}$. The clusters have half-light radii
of 7.5 -- 14 pc, and colour-magnitude diagrams that are consistent
with a wide range of ages (2 -- 10 Gyr) and metallicities (Z = 0.0001
-- 0.004). These clusters are very similar to the extended clusters
found in M31 \citep{Huxoretal05} and M33 \citep{Stonkuteetal08}, with
a couple of the clusters (SC1 and SC4) being very distant from NGC
6822 itself.  \citet{Hwangetal11} also noted that the extended
clusters project on a line that is consistent with the major axis of
the old stellar halo.

Drawing partly on the data from the Hubble clusters,
\citet{Straderetal03} suggest a view of NGC 6822 in which there has
been a relatively constant star formation rate over time, with
occasional stochastic outbursts that result in the formation of star
clusters. A similar scenario has also been outlined by
\citet{ColucciBernstein11}.

\section{The Search for NGC 6822 Clusters}

\begin{figure}
 \centering
 \includegraphics[angle=90,width=85mm]{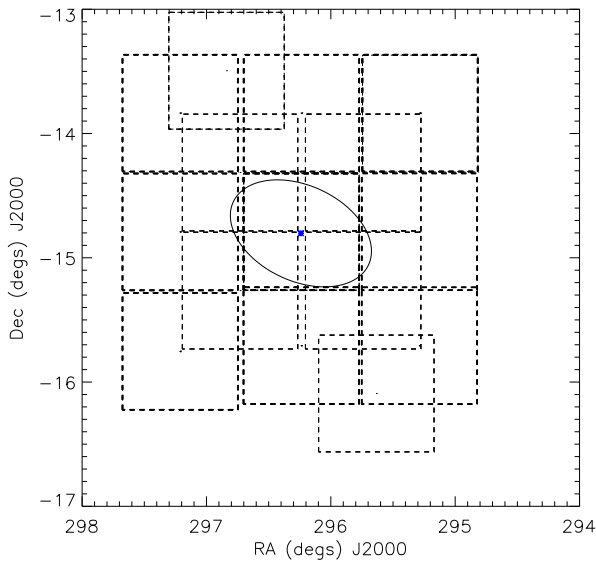}
 \vspace{2pt}
 \caption{Locations of the CFHT/Megacam fields studied. The centre of
   NGC 6822 is represented by the small solid square, and the ellipse
   is the same as that in Figure \ref{Fi:dss}}\label{Fi:fields}
\end{figure}

An initial study of the archives found that NGC 6822 had considerable
and contiguous coverage in CFHT/Megacam imaging (see Figure
\ref{Fi:fields}). This is a wide-field camera at the
Canada-France-Hawaii Telescope (CFHT) with a 1$^{\circ}$ $\times$
1$^{\circ}$ field of view and a pixel scale of 0$\arcsec$.187. We
naturally use the same imaging as that of \citet{Hwangetal11}, but
also include additional fields that extended the coverage and fill the
gaps between the CCDs in their survey. In total, we searched 15
CFHT/Megacam fields from the programs 2003BK03, 2004AC02, 2004AQ98,
and 2005AK08, with observations taken over the period August
2003--August 2006. Science exposures ranged from 660 -- 1200 seconds
in the g-band, 360 -- 1000 seconds in the r-band, and 150 -- 460
seconds in the i-band.

The images were visually inspected since star clusters at NGC 6822's
distance are easily resolved in CFHT/Megacam imaging, and indeed this
is the optimal way to identify any additional examples of the extended
clusters. We are only concerned with the outer regions of NGC 6822,
and do not study the main body of the galaxy where many young clusters
have already been documented \citep{Krienkeetal04}.

We also examined archival Subaru/Suprime-Cam imaging of NGC 6822. This
instrument has a $\sim$ 0.5$^{\circ}$ $\times$ 0.5$^{\circ}$ field of
view and a pixel scale of 0$\arcsec$.20, and the imaging was mainly
concentrated on the inner regions of the galaxy. We utilised only
those images for which the exposure was greater than 200 seconds.
Those available in the archive were obtained in B,V,R or I-band
filters, and were taken for proposals o01422, o00005, o02419,o03147,
o99005, o04151, and o05226.  Although these pointings did not extend
much beyond the main body of NGC 6822, many of the images were deeper
than those from CFHT/Megacam and they proved useful to confirm, or
otherwise, candidate clusters found in the CFHT/Megacam imaging.

\section{ The New Clusters}

\begin{figure}
\begin{center}$
\begin{array}{cc}
\includegraphics[angle=0,width=40mm]{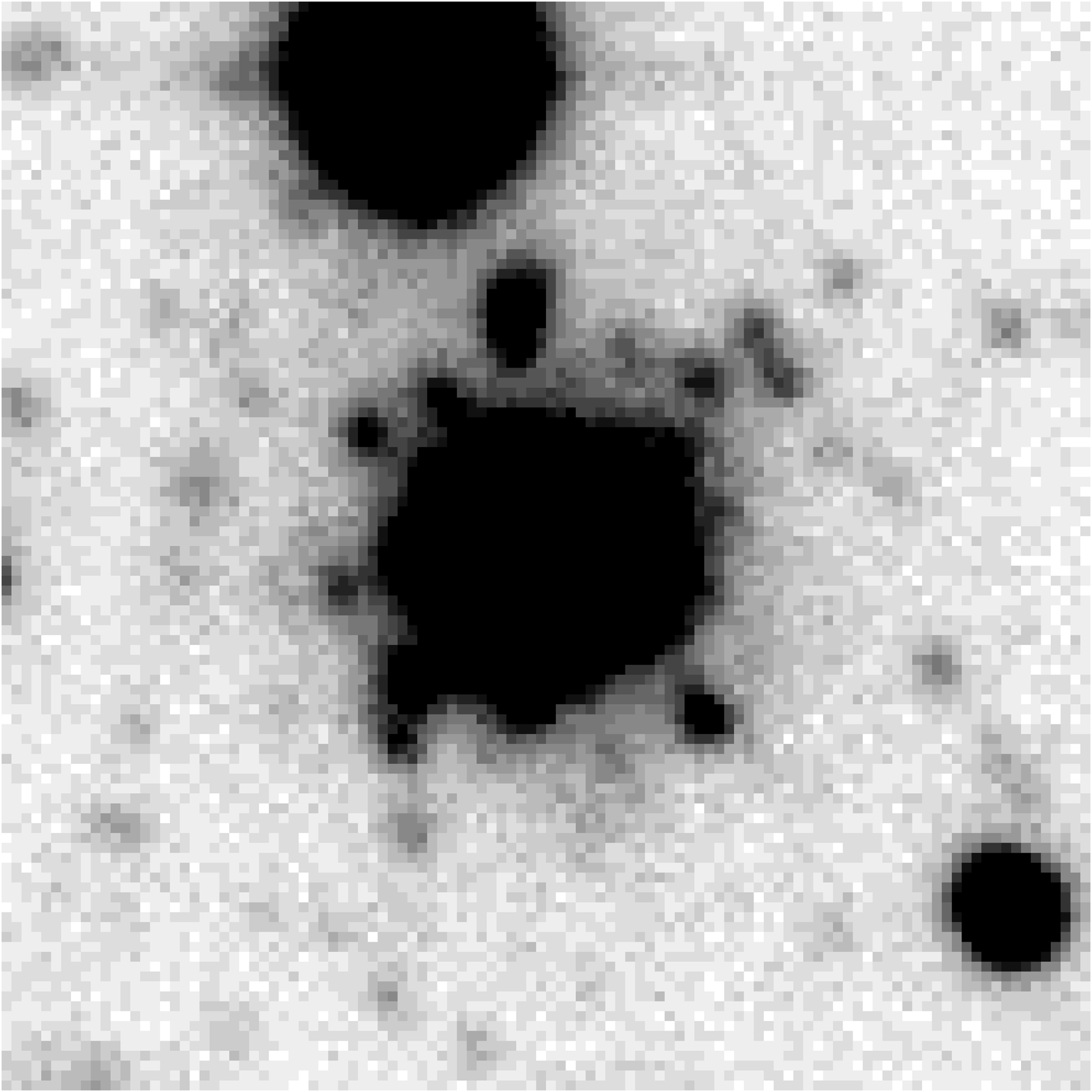} &
\includegraphics[angle=0,width=40mm]{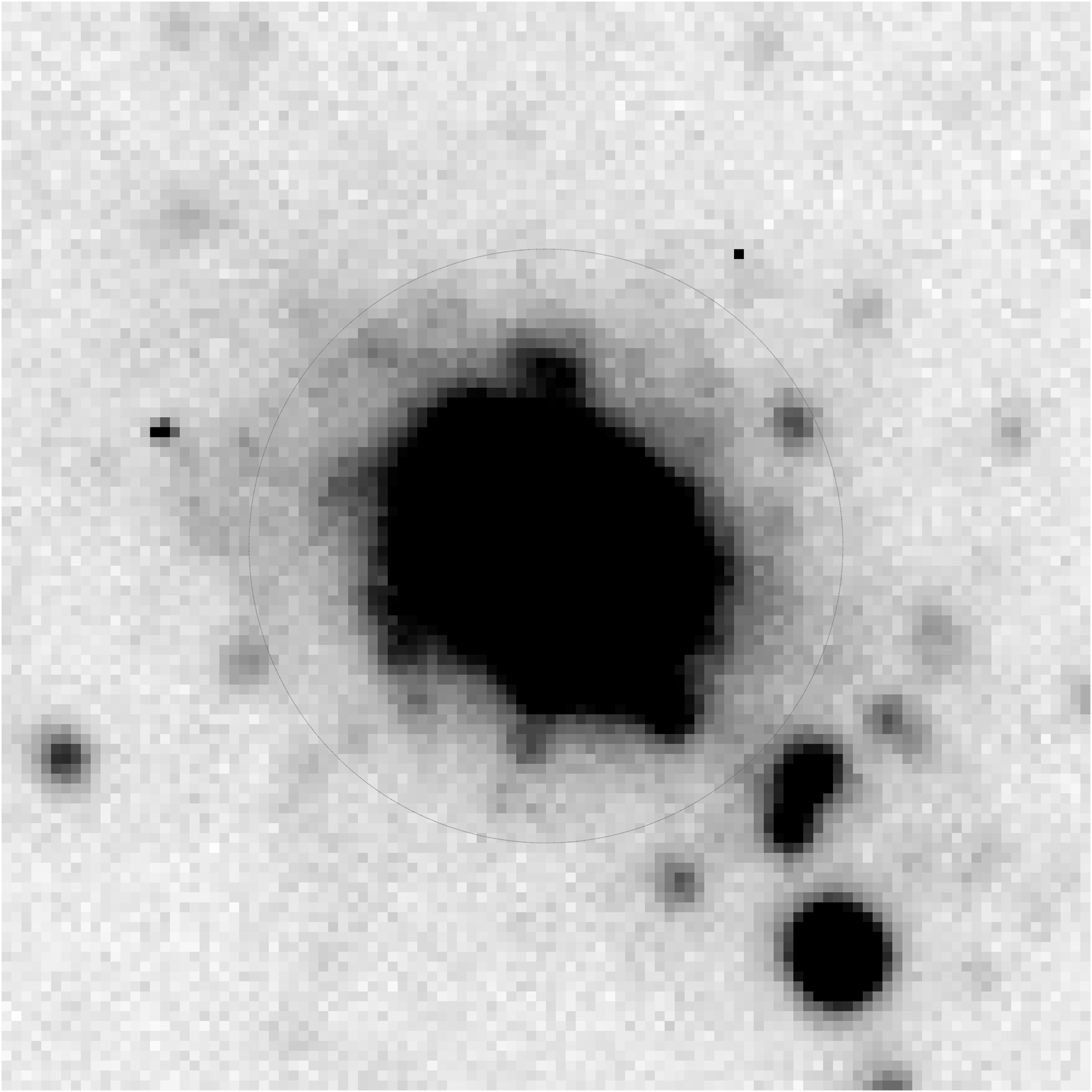} \\
\end{array}$
\end{center}
\caption{The two new compact GCs, SC6 (left) and SC7 (right) from
  CFHT/Megacam i-band imaging. Each image is 20 by 20 arcsec. North is
  up and East is to the left. The circle overlaid on SC7 has a radius
  of 30 pixels (c.f. Figure \ref{Fi:A_pa_ellipticity}).
}\label{Fi:image_AB}
\end{figure}

\begin{figure}
\begin{center}$
\begin{array}{cc}
\includegraphics[angle=0,width=40mm]{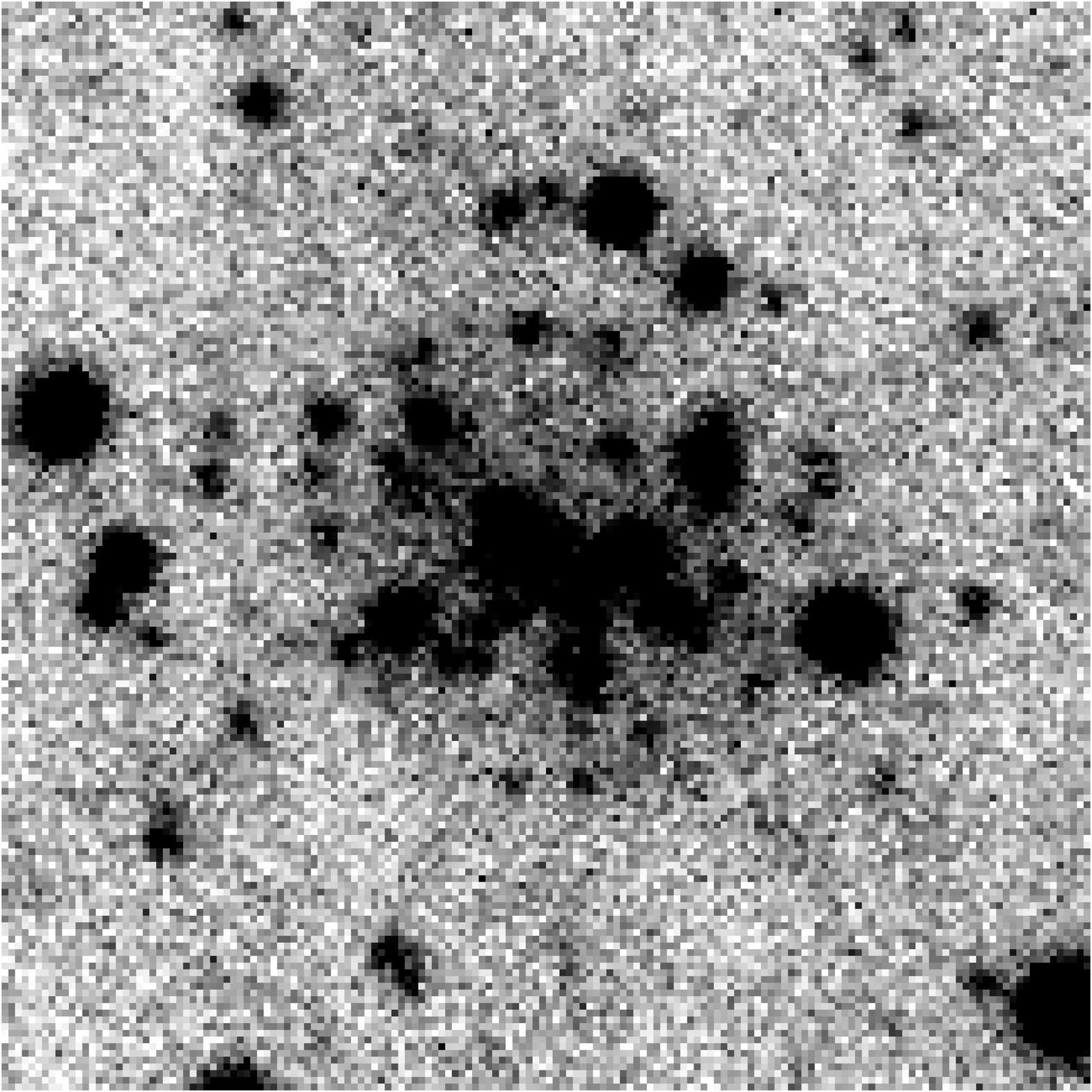} &
\includegraphics[angle=0,width=40mm]{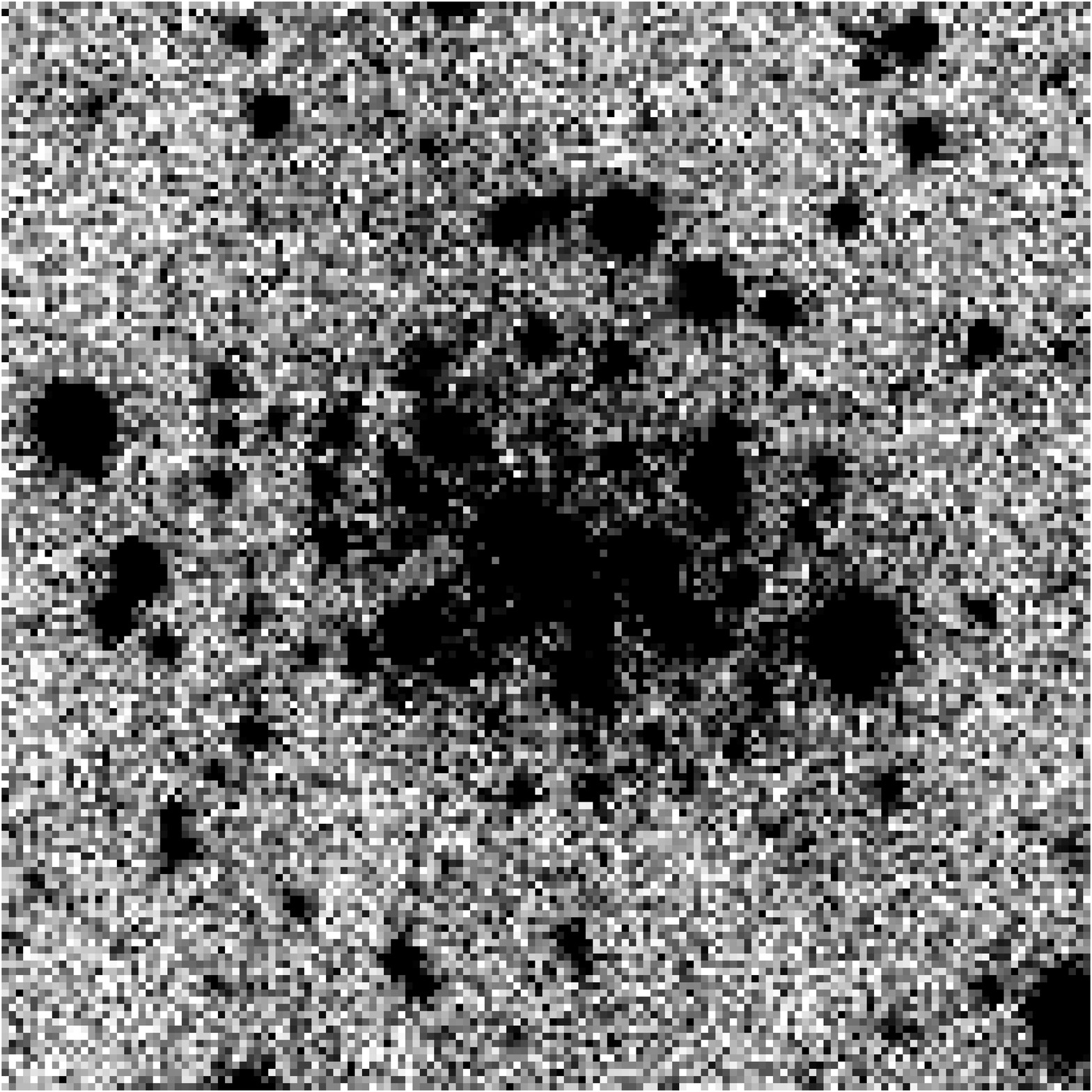}
\end{array}$
\end{center}
\caption{The new extended GC, SC5, from r-band CFHT/Megacam (left) and
  Cousins-I-band Subaru/Suprime-Cam (right) imaging, stretched to
  highlight the faint cluster. Each image is 30 by 30 arcsec. North is
  up and East is to the left.  }\label{Fi:image_D}
\end{figure}

The search for new GCs found a total of three new clusters, in
addition to rediscovering all those of \citet{Hwangetal11}. Two are
luminous compact classical clusters, and one is very faint and appears
extended in form. We continue the naming convention used by
\citet{Hwangetal11}, and denote them as SC5, SC6 and SC7 (in order of
right ascension). The coordinates of these objects and their projected
distance from the centre of NGC 6822 are listed in Table
\ref{tab:positions}. The two luminous clusters (SC6 and SC7) are clear
examples of GCs (see Figure \ref{Fi:image_AB}). The new faint cluster
(SC5) is, in contrast, much more diffuse.  Although barely detected in
a single exposure from the CFHT archive, SC5 can be seen more clearly
in a stacked image available through the CFHT archive (Figure
\ref{Fi:image_D}, left panel), and also in a deep archival
Subaru/Suprime-Cam image (Figure \ref{Fi:image_D}, right panel). SC5
resolves into stars while SC6 and SC7 only do so in their peripheries.

\begin{table}
 \centering
\begin{minipage}{80mm}
  \caption{Locations of the new clusters, and their projected
    distances (R$_{proj}$) from the centre of NGC 6822 (RA = 19$^{h}$
    44$^{m}$ 57.7$^{s}$, Dec = --14$^{\circ}$ 48$\arcmin$
    12$\arcsec$).}\label{tab:positions}
\vspace{2pt}
\begin{tabular}{@{}lllc@{}}
\hline
 ID	& RA(J2000) & Dec(J2000) & R$_{proj}$ (kpc) \\
\hline
 SC5		& 19$^{h}$ 43$^{m}$ 42.30$^{s}$   	& --14$^{\circ}$ 41$\arcmin$ 59.7$\arcsec$		& 2.6	 	\\ 
 SC6		& 19$^{h}$ 45$^{m}$ 37.02$^{s}$   	& --14$^{\circ}$ 41$\arcmin$ 10.8$\arcsec$ 		& 1.6		\\ 
 SC7		& 19$^{h}$ 46$^{m}$ 00.85$^{s}$ 	& --14$^{\circ}$ 32$\arcmin$ 35.4$\arcsec$ 		&  3.0	\\ 

\hline
\end{tabular}
\end{minipage}
\end{table}

\subsection{Integrated Photometry}

Integrated photometry was undertaken for the two most luminous new
clusters, using the archival imaging data available. The results are
reported in CFHT/Megacam filter magnitudes (which are similar but not
identical to standard Sloan filters) in Table \ref{tab:properties}.

In our photometry we used large apertures that enclose the full extent
of the cluster for the total magnitudes. As there is no evidence that
GCs have strong colour gradients, we employed smaller apertures to
obtain more reliable colours. Photometric calibration of the CFHT data
was undertaken using the magnitudes derived for the one pointing taken
in photometric conditions, and cross-calibrating the other data using
stars common to both.

Photometry for the brightest cluster, SC7, proved problematic. In the
archival CFHT/MegaCam data, the cluster is saturated in the g- and
r-bands and photometry can only be undertaken in the i-band. However,
shorter exposures of cluster SC7 were also in the CFHT archive, taken
for the purposes of photometric calibration.  In these exposures, SC7
unfortunately lands on the edge of a CCD making measurements of the
full cluster impossible. Hence, we use the central region of the
shorter exposures to obtain the colours using an aperture radius of
1.5$\arcsec$.  We then estimate the total magnitudes by using an
aperture of radius 6$\arcsec$ on the long i-band image and applying
the colour measurements to obtain total g- and r-band magnitudes. For
cluster SC6, no such problem arose: the apertures employed for the
deriving the color and total magnitude had radii of 2$\arcsec$ and
4.7$\arcsec$ respectively.

The very faint cluster SC5 was also difficult to photometer. This
object is visible in a long CFHT/Megacam r-band stack (11000 seconds)
but the g and i-band data, even when stacked, are too shallow to
detect the cluster. The r-band stack was photometered with an aperture
radius of 10$\arcsec$. SC5 was also found in archival Subaru data,
confirming its status as a cluster.

Photometry in the CFHT/Megacam filter set was also converted to
Johnson-Cousins V and I for SC6 and SC7. This was achieved by using
the colour transform equations given on the SDSS
web-pages\footnote{http://www.sdss.org/dr4/algorithms/sdssUBVRITransform.html}.
This was not possible for SC5 as we require photometry in more than
the one filter for the transform equations.

Extinction is known to be a major issue with NGC 6822 due to its low
Galactic latitude.  \citet{Battinellietal06} use the stellar
population of NGC 6822 to estimate the foreground reddening across the
area discussed in this paper, and find it is not only significant, but
also patchy. Specifically, E(B-V) ranges from 0.19 to 0.30 (their
Figure 2).  We correct for this using the extinction maps --
interpolated to the position of the new clusters -- and relative
extinction for the Sloan band-passes from \citet{Schlegeletal98}.  NGC
6822 also has internal reddening and \citet{Masseyetal95} find values
of up to E(B-V) of 0.45 mags in the centre of the galaxy. However the
new clusters lie far from the centre of NGC 6822 and should be
minimally impacted by internal reddening.  We note, however, that
patchy Galactic extinction may limit the accuracy our final
photometry.

\begin{table}
 \centering
\begin{minipage}{85mm}
  \caption{Photometric properties of the new clusters. Extinction
    corrections use values estimated from the extinction map of
    \citet{Schlegeletal98}. The g, r and i-magnitudes are in the
    CFHT/Megacam filter system. Photometric errors on the colours
    (derived for a inner aperture for SC6 and SC7 -- see text) are
    estimated at $\pm$0.01 magnitudes. The major source of error for
    the total magnitudes is the uncertainty of the memberships of
    cluster stars within the aperture, which are estimated at
    $\pm$0.03 magnitudes.}\label{tab:properties}
 \vspace{2pt}
\begin{tabular}{@{}lccccccc@{}}
\hline
 ID 		& g$_{0}$ & r$_{0}$  &  i$_{0}$  & V$_{0}$ & (V-I)$_{0}$ & M${_{V}}_{0}$ & E(B-V)  \\
\hline
 SC5		& --   		& 19.43	& --		& -- 		& --	 	& --	 	& 0.219		\\ 
 SC6		& 15.55   	&  15.01 	& 14.79		& 15.28	& 0.84	& --8.09 	&  0.190	\\ 
 SC7		& 15.02   	&  14.43	& 14.10	& 14.77 	& 1.05	& --8.60 	&  0.207	\\ 

\hline
\end{tabular}
\end{minipage}
\end{table}

\begin{figure}
\begin{center}
 \includegraphics[angle=0,width=85mm]{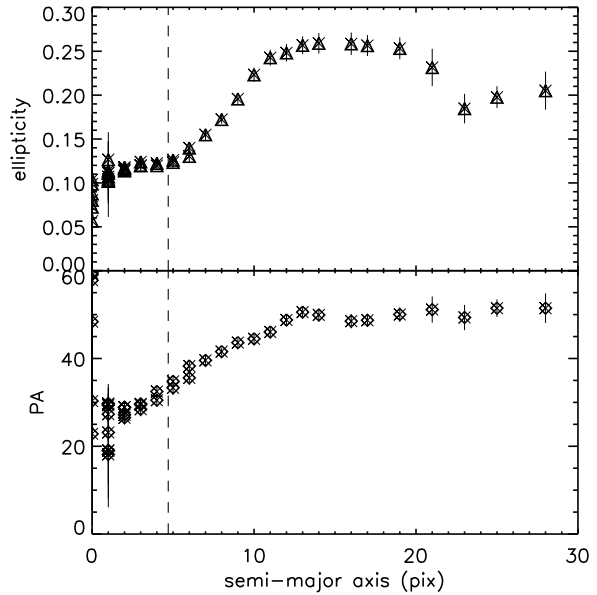}
 \vspace{1pt}
 \caption{Ellipticity and PA as derived from IRAF/ELLIPSE for SC7.
   The PA is $\sim$50$\deg$ over radii of $\sim$10--30 pixels (c.f.
   circle on left panel of Fig.\ref{Fi:image_AB}). The ellipticity has
   a value of $\sim$0.25 over a large range of radii. The FWHM of the
   image is 4.7 pixels (dashed vertical
   line).}\label{Fi:A_pa_ellipticity}
 \end{center}
\end{figure}

\subsection{Ellipticity of SC7}

\begin{figure}
\begin{center}
 \includegraphics[angle=0,width=85mm]{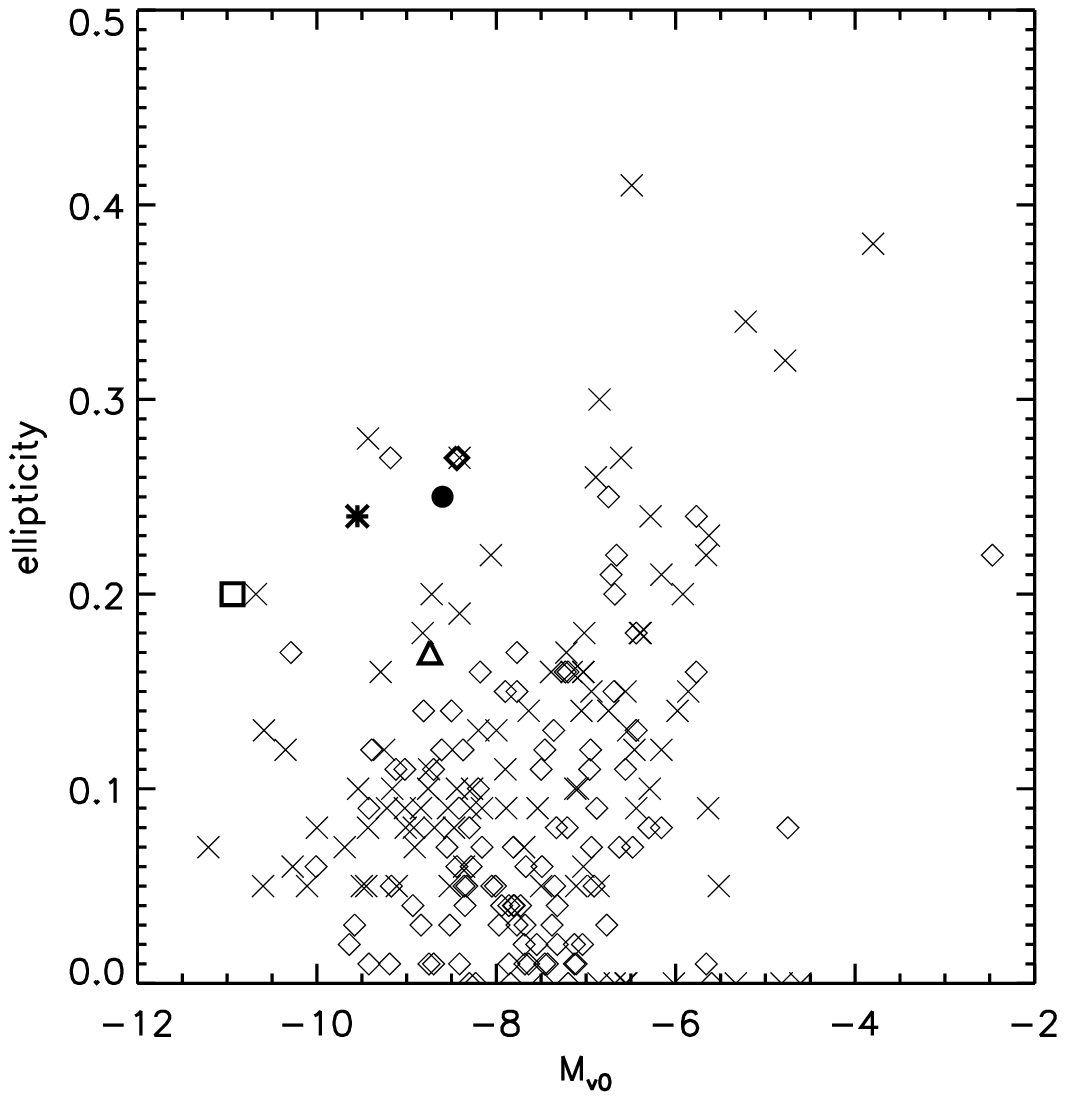}
 \vspace{1pt}
 \caption{The location of SC7 (filled circle) in a plot of
   M${_V}$-ellipticity. Also shown are MW GCs (diamonds) taken from
   the McMaster MW GC catalogue \citep{Harris96}, M31 clusters taken
   from \citet{Barmbyetal07} (crosses) and G1 \citep{Maetal07}
   (square), WLM-1 \citep{Stephensetal06} (triangle), cluster 77 from
   \citet{Annibalietal11} (asterisk,) and NGC 121 \citep{Glattetal09}
   (thick diamond), where the magnitude for NGC 121 is the mean of the
   values they derive for a King and EFF fit to the profile. The most
   elliptical GC in the MW is M19, but this is known to be a result of
   differential reddening \citep{vandenBergh08}.
 }\label{Fi:ellipticities}
 \end{center}
\end{figure}

Visual inspection reveals that cluster SC7 is significantly elongated.
We used IRAF/ELLIPSE\footnote{IRAF is distributed by the National
  Optical Astronomy Observatories, which are operated by the
  Association of Universities for Research in Astronomy, Inc., under
  cooperative agreement with the National Science Foundation.} to
derive the ellipticity and position angle (PA) of the major axis of
SC7 using a fixed centre and the results are shown in Figure
\ref{Fi:A_pa_ellipticity}. The PA beyond $\sim$ 12 pixels is
50$^{\circ}$ and the ellipticity has a value of $\sim$0.25 over the
main body of the cluster. This high ellipticity is unusual for a GC
and makes SC7 a clear outlier in a plot of M${_{V}}_{0}$ versus
ellipticity (Figure \ref{Fi:ellipticities}).
 
\section{ Discussion and Summary}

The new clusters reported here substantially increase the number of
classical GCs found in NGC 6822.  If all the new massive clusters
(excluding SC5, which is too faint to be found in comparable studies
of local dwarfs) prove to be genuinely ``old" GCs, then the four
clusters in \citet{Hwangetal11} and the two in this work would
increase the specific frequency of NGC 6822 to $S_{N}$ $\sim$ 7,
comparable to the newly-enlarged GC systems of NGC 147 and NGC 185
(Veljanoski et al, in prep.). This value is also consistent with
values found for dwarf irregulars in the Virgo and Fornax galaxy
clusters \citep{Sethetal04}.

The cluster SC7 is of relatively high luminosity and SC6 is almost as
luminous, with M${_{V}}_{0}$ $\sim$ --8 mags. The GC systems of M31
and the Milky Way have median values of M$_{V}$ are --7.9 and -7.3
respectively \citep{Huxoretal11}, so both SC6 and SC7 are brighter
than the turnover of the globular cluster luminosity function for
these galaxies. Previously, in \citet{Mackeyetal07} and
\citet{Huxoretal11}, we found that M31 possesses a number of luminous
GCs in its outer stellar halo, for which no counterparts exist in the
Milky Way (excepting the very unusual cluster NGC 2419). If, as seems
likely, the accretion history of M31 was different from that of the
MW, we may have a natural source of M31's luminous halo GCs in the
accretion of systems such as NGC 6822.  However such events would have
had to happen at an early epoch since there is no evidence for young
populations - which dominate in galaxies like NGC 6822 - in the M31
halo today.

The origin of high ellipticities in GCs, such as that of SC7, has been
the source of some debate. \citet{Kontizasetal90} find that the
ellipticity for young SMC star clusters is greater than for the
clusters in the somewhat more massive LMC, and similar results lead
\citet{Georgievetal08} to argue that the tidal field of the host
galaxy is likely to be an important factor in determining cluster
ellipticity. A scenario in which the SC7's ellipticity is a
consequence of it being formed in a dwarf galaxy host is also
consistent with the presence of the extended clusters in NGC 6822.
Indeed, \citet{HurleyMackey10} argue that the formation and survival
of extended clusters is a consequence of the more benign tidal fields
in dwarf galaxies and the outer regions of massive galaxies. The
origin of the high natal ellipticity, which a low tidal field
preserves, may arise from a number of sources: rotation, galactic
tides or anisotropy in the velocity dispersion between the major and
minor axes. The latter was found to be the best explanation for the
high ellipticity of the only GC known thus far in the dwarf galaxy WLM
\citep{Stephensetal06}.
 
It should be noted that an alternative scenario for the formation of
luminous, elliptical GCs has also been proposed.  In a study the star
cluster system of the Magellanic-type starburst galaxy NGC 4449,
\citet{Annibalietal11} found that the brighter clusters tend to be
more elliptical. One of their clusters (cluster 77) is old, massive
and highly elliptical (see figure \ref{Fi:ellipticities}), leading
them to suggest that it may be the nucleus of a satellite galaxy that
is currently being stripped \citep{Annibalietal12}.  A similar picture
has also been proposed for G1, the most luminous GC in
M31\citep{Meylanetal01}, and the anomalous Galactic GC $\omega$ Cen
\citep[e.g.][]{Romanoetal07}.

One last notable aspect of the newly discovered GCs is that they lie
in the linear arrangement noted by \citet{Hwangetal11}. As we have
surveyed the full area in Figure \ref{Fi:fields}, this distribution
cannot be a result of incomplete areal coverage.  Such a disk
alignment would not be unusual -- the cluster population of the LMC
exhibits disk-kinematics \citep{Schommeretal92,Grocholskietal09} --
but it would raise new questions about the formation of NGC 6822.  If
the GC system is found to exhibit disk-like kinematics, it might be
hard to reconcile with a scenario where the galaxy formed via the
merger of two similar mass gas-rich dwarfs \citep{Bekki08}. We have
spectroscopic data for SC6 and SC7, and are currently obtaining data
for other clusters in NGC 6822 to study the kinematics of the cluster
system and address this question.

To summarise, we have presented the discovery of three new star
clusters in the outskirts of NGC 6822 based on searches conducted with
archival datasets. Two of these objects are massive compact GCs, very
distinct from the extended clusters found by \citet{Hwangetal11}. The
third is a very low-luminosity diffuse cluster. We have measured
integrated photometry for these objects, but additional
characterisation (e.g. structural parameters, stellar populations)
will require deep high resolution data.  SC6 and SC7 are so compact
that R$_{h}$ is comparable to the FWHM in the data presented here.
One of the clusters, SC7, is highly elliptical which we speculate
could be due to the low tidal field it has experienced in the outer
regions of a dwarf galaxy.

We note in closing that it is remarkable that SC6 and SC7 were not
discovered earlier. These are high luminosity GCs in a Local Group
galaxy that has been studied very extensively.  This underscores yet
again how the outer regions of galaxies have the ability to surprise
and provide important clues about their histories.

\section*{Acknowledgments}

This work was partly supported by Sonderforschungsbereich SFB 881 "The
Milky Way System" of the German Research Foundation (DFG). ADM is
grateful for support by an Australian Research Fellowship (DP1093431)
from the Australian Research Council.  AHP, AMNF and ADM acknowledge
support by a Marie Curie Excellence Grant from the European Commission
under contract MCEXT-CT-2005-025869 during the early stages of this
work.  We also would like to thank Erwin de Blok for kindly providing
the data for the HI map used in figure \ref{Fi:dss}.

Based on observations obtained with CFHT/MegaCam, a joint project of
CFHT and CEA/DAPNIA, at the Canada-France-Hawaii Telescope which is
operated by the National Research Council (NRC) of Canada, the
Institute National des Sciences de l'Univers of the Centre National de
la Recherche Scientifique of France, and the University of Hawaii.
The United Kingdom Infrared Telescope is operated by the Joint
Astronomy Centre on behalf of the Science and Technology Facilities
Council of the U.K.

\end{document}